\documentstyle[preprint,aps]{revtex}
\begin{document}
\draft

\title{Tunnelling  studies of two-dimensional states in semiconductors
with inverted band structure: Spin-orbit splitting, resonant
broadening}

\author{G.M.Minkov, A.V.Germanenko,V.A.Larionova,O.E.Rut}
\address{Institute of Physics and Applied Mathematics, Ural
University, Ekaterinburg 620083, Russia}
\date{\today}
\maketitle
\begin{abstract}
The results of tunnelling studies of the energy spectrum of
two-dimensional (2D) states in a surface quantum well in a
semiconductor with inverted band structure are presented. The
energy dependence of quasimomentum of the 2D states over a wide
energy range is obtained from the analysis of tunnelling
conductivity oscillations in a quantizing magnetic field. 
The spin-orbit splitting of the energy spectrum of 2D
states, due to inversion asymmetry of the surface quantum well,
and the broadening of 2D states at the energies, when they
are in resonance with the heavy hole valence band, are
investigated in 
structures with different strength of the surface quantum well.
A quantitative analysis is carried out within the framework of
the Kane model of the energy spectrum. The theoretical results
are in good agreement with the tunnelling spectroscopy data. 

\end{abstract}

\narrowtext
\section{Introduction}
\label{sec:intro}

     A specific feature of a semiconductor with inverted energy 
band structure is the absence of an energy gap between conduction 
and valence bands. 
These bands are the two branches of the  
fourfold $\Gamma_{8}$ band. The states of
$\Gamma_{8}$ band are  
classified by the projection of the angular momentum J=3/2 onto 
the direction of quasimomentum  $k$. The states with a projection
of $\pm 1/2$ are conduction band states, hereafter referred to as 
the spin states of an electron.  The $\Gamma_{6}$  band, which 
is a conduction band in ordinary semiconductors
$\text{A}^{3}\text{B}^{5}$ and $\text{A}^{2}\text{B}^{6}$ with 
an open gap, is the light hole band, and lies below the degeneracy 
point for the energy difference $E_{g}
=E^{\Gamma_{6}}-E^{\Gamma_{8}}$. 

     Such peculiarities of the band structure of 
inverted semiconductors lead to some special features 
of the energy spectrum of spatially confined systems, based on 
these semiconductors. Thus, it was predicted theoretically 
\cite{1} that so called interface 2D states can exist 
near the boundary of an inverted semiconductor even without an 
attractive electrostatic potential. It was shown in a  
number of  theoretical  and 
experimental articles \cite{2,3,4,5,6,7,8} that these states
play a key part in  
forming the energy spectrum of 2D states in heterostructures and 
superlattices based on inverted semiconductors. 

     The absence of a forbidden gap in the inverted semiconductors 
lead to the fact that at negative energies (hereafter we measure 
energy from the degeneracy point of $\Gamma_{8}$ band in the 
volume) 
the 2D states in the surface quantum well are  resonant
with the heavy hole valence band and, therefore, broadening
should be present in this energy range. Such effects have been 
discussed for 
narrow gap semiconductors. However in these materials, 2D electron 
states at energies less than the energy of the top of the valence band 
are separated from the heavy and light hole states by a barrier and 
the width of this barrier is proportional to the forbidden gap. 
So, the effect of the resonanse broadening in ordinary 
semiconductors should differ from that in gapless 
semiconductors.

     Of special interest are 2D systems in asymmetric quantum 
wells. The lack of inversion symmetry in this case leads to the 
spin splitting of the 2D subbands at $k\ne0$ due to spin-orbit 
coupling even without a magnetic field. This phenomenon for 2D 
electron states has been studied extensively in semiconductors
with $E_{g}>0$.  
\cite{9,10,11,12,13,14} In such materials the conduction band 
is the twofold degenerate $\Gamma_{6}$ band and the spin-orbit 
interaction can be taken into account within perturbation 
theory. \cite{14} As mentioned above, in inverted 
semiconductors the conduction band is one of branches of 
$\Gamma_{8}$ 
band and the spin-orbit coupling has been considered from the 
onset.\cite{15}

     This work is devoted to the investigation of the energy 
spectrum of 2D states in the surface quantum well in the 
inverted semiconductor
$\text{{Hg}}_{1-x}\text{Cd}_{x}\text{Te}$  using tunnelling
spectroscopy in a 
quantizing magnetic field. This method was used for the first
time in an investigation of the 2D states in InAs surface  
accumulation layer by Tsui in Ref.\ \onlinecite {16}. Since
then,  2D states in a  
large variety of semiconductors, e.g. GaAs \cite{17}, InSb, 
\cite{17a} and InAs, \cite{18} were studied by tunnelling
spectroscopy in a magnetic  
field. However there are only a few  tunnelling experiments on 
structures based on inverted semiconductors. \cite{8,19}

     Unlike traditional methods (galvanomagnetic phenomena, 
volt-capacitance experiments), which give  information about 
the carriers at the Fermi energy, this method allows one to obtain 
information about the energy dependence of quasimomentum of 
carriers over a wide range of energies both for empty and occupied 
states. In this work we have also used a refinement 
of tunnelling spectroscopy which allowed us 
to investigate  2D states at energies, which are given by the 
applied bias, for different quantum well
strengths.

     In the present article in contrast to  previous
publications \cite{8,19} we report on 
results obtained on tunnel structures with a larger strength 
of the surface potential well. The preparation of such tunnel 
structures has been made possible by the use of Yb as a metal 
electrode, which has a low work function. 
This allows us to observe both 
spin branches, which are split by the spin-orbit interaction,
and 2D states at  large negative energies where they are in
resonance with the  valence band.

     The article is organized as follows. In Section
\ref{sec:tun} some features of the tunnelling spectroscopy of
2D states in a quantizing  
magnetic field, are considered. The experimental details and 
innovations in the tunnelling spectroscopy method are given in 
Section \ref{sec:det}. In Sec. \ref{sec:exp}, experimental
results obtained for the  
tunnel structures with different strengths of the surface 
potential well are presented and analysed. Section \ref{sec:theor} is 
devoted 
to the theoretical description of the energy spectrum of 2D 
states in a surface quantum well of an inverted semiconductor.
The basic equations, used in the  
calculations of resonant 2D states, are spelled out. A 
comparison between experiment and theory is discussed in Section 
\ref{sec:disc}, and, finally, conclusions are made in Section 
\ref{sec:conc}.

\section{Oscillations of the tunnelling conductivity in a magnetic 
field}
\label{sec:tun}

     In principle, the bias dependence of tunnelling conductivity of a
metal-insulator-semiconductor structure contains information on 
the energy spectrum of both 3D  and 2D states which may exist in 
the surface quantum well of the semiconductor. The investigations 
of tunnelling conductivity oscillations in a magnetic field permit 
one to obtain the most reliable information on the energy 
spectrum, because in this case it is possible to determine directly 
the positions of the Landau levels and energy intervals between 
them. 

     Different types of oscillations of the tunnelling 
conductivity may occur in the  metal-insulator-semiconductor 
structure with 2D states localized in the surface quantum well
(Fig.\ \ref{fig1}).   
The tunnelling current in such a structure is the sum of the current 
due to tunnelling to 2D ($j_{2D}$) and to 3D ($j_{3D}$) states. The 
oscillations of $\sigma_{3D}\equiv dj_{3D}/dV$ versus magnetic 
field and bias 
have been considered earlier. \cite{19a,20,21}
The maxima in $\sigma_{3D}$   occur when the 
metal Fermi level is aligned with one of the bulk Landau levels
of the semiconductor. These oscillations are periodic in the
reciprocal magnetic field with the period
\begin{equation}
\Delta(E_{F}+eV)=\frac{2 \pi e}{c\hbar S(E_{F}+eV)} 
\label{eq1}
\end{equation}
where $E_{F}$ is the semiconductor Fermi energy and 
$S(E_{F}+eV)$
is the extreme cross-section of the iso-energy surface  
at the energy $E_{F}+eV$. In semiconductors with an isotropic 
spectrum 
$S(E)=\pi k^2(E)$. This  gives one the opportunity 
to determine the energy dependence of the quasimomentum for 
the bulk states over a wide energy range.\cite{20,21} 

     The tunnelling current to 2D states at bias $V$, magnetic
field $B$, and temperature $T=0$ is
\begin{equation}
j_{2D}(V)=\int_{E_{F}}^{E_{F}+eV}g_{2D}(E-E_{0},B)D(E,V)dE 
\label{eq2}
\end{equation}
where $g_{2D}(E-E_{0},B)$, $E_{0}$, $D(E,V)$ stand for the density 
of
2D states, energy  
of the bottom of 2D subband and the barrier tunnelling transparency, 
respectively.  An applied voltage not only shifts the Fermi 
quasilevels of the metal and semiconductor relative to each other 
but also changes the surface quantum well (and therefore $E_{0}$) 
and barrier 
transparency as well. Thus the tunnelling conductivity for 
tunnelling to 2D states is
\widetext
\begin{eqnarray}
\sigma_{2D}\equiv\frac{dj_{2D}}{dV}&=&g_{2D}(E-
E_{0},B)D(E,V)|_{E_F+eV}+
\nonumber \\
&&+\int^{E_F+eV}_{E_F}\left(\frac{dg_{2D}(E-
E_{0},B)}{dV}D(E,V)+
g_{2D}(E-E_{0},B)\frac{dD(E,V)}{dV}\right)dE
\label{eq3}
\end{eqnarray}
To a first approximation, the variation of $g_{2D}(E-E_{0},B)$ with 
bias is 
due to changes in $E_{0}$, and relationship between $E_{0}$  and 
$V$ is 
linear, i.e. $E_{0}(V)=E_{0}(0)+\alpha V$ and
$g_{2D}(E-E_{0},B)=g_{2D}(E-E_{0}(V),B)$. This 
results in the expression 
\begin{eqnarray}
\sigma_{2D}&=&D(E,V)g_{2D}(E-E_{0}(V),B)|_{E_F+eV}+
\alpha\overline{D}(V)g_{2D}(E-E_{0}(V),B)|_{E_F+eV}-
\nonumber \\
&&-\alpha\overline{D}(V)g_{2D}(E-E_{0}(V),B)|_{E_F}+
\int_{E_F}^{E_F+eV}g_{2D}(E-E_{0}(V),B)\frac{dD(E,V)}{dV}dE
\label{eq4}
\end{eqnarray}
\narrowtext 
where $\overline{D}(V)$ is the mean value of $D(E,V)$. A magnetic 
field $B
\parallel n$ ($n$ is the  
normal to the surface) quantizes the energy spectrum of 2D states 
and $g_{2D}$ becomes an oscillatory function of $B$ and $E$. One 
can see 
from expression (\ref{eq4}) that two types of conductivity oscillation
should occur in a magnetic field.

     (i) In the first type, oscillations arise whenever a Landau 
level of 2D states is aligned with the energy $E_{F}+eV$ (the first 
and the 
second terms in Eq.\ (\ref{eq4}). If the surface potential depends 
only 
slightly on $V$, then the positions of these oscillations allow
one to immediately determine the energy spectrum of the 2D
states by the same procedure as for 3D states (see Eq.\ (\ref{eq1})). 
As
$B\rightarrow 0$ the fan chart of these oscillations is
extrapolated  to the
energy of the bottom of the 2D subband.

     (ii) In the second type, oscillations appear whenever a 
Landau level of the 2D states is aligned  with the energy $E_{F}$ 
(the 
third term in Eq.\ (\ref{eq4})). As $B\rightarrow 0$ the fan chart
of these oscillations is  
extrapolated to the bias at which the 2D carriers (but not states) 
disappear, i.e. to the bias for which the bottom of the 2D subband 
becomes higher, than the semiconductor Fermi level. The amplitude 
of these oscillations is proportional to the rate of variation
of $E_0$ with bias ($\alpha$). When $\alpha=0$,
or $E_{0}>E_{F}$  these oscillations are absent and  
only oscillations of the first type will be observed.

     We do not concern oscillations of the forth term in Eq.\
(\ref{eq4}), because an oscillating function is an integrand, 
and, therefore, the amplitude of oscillations of this term is small.
 
     Up to now we have not considered  possible oscillations
of $D(E,V)$ in a magnetic field. The effects of these oscillations
might be dominant in tunnel structures with a monocrystal
barrier. For such a barrier the decay  
constant of the wave function in an insulator is rigidly determined 
by the energy. Therefore, small oscillations of the surface
potential $\varphi_{s}$ (Fig.\ \ref{fig1}) give rise to 
significant oscillations of $D(E,V)$. Oscillations of
$\varphi_{s}$ as well as oscillations of the 
third term in Eq.\ (\ref{eq4})   occur whenever a Landau 
level of the 2D states is aligned with the energy $E_{F}$. \cite{16} 
An 
analysis of the expression (\ref{eq3}) shows that the second
type of oscillations  
will be predominantly observed in such structures. We believe 
that the oscillations of this kind  were observed in 
HgTe/HgCdTe heterostructures. \cite{22}

\section{Experimental details}
\label{sec:det}
     The differential conductivity $\sigma_{d} =dj/dV$ as a
function of bias and 
magnetic field in
metal-insulator-$\text{Hg}_{\text{1-
x}}\text{Cd}_{\text{x}}\text{Te}$
 ($0.08<\text{x}<0.13$) 
structures was investigated in magnetic fields up to 6 T at the 
temperature of 4.2 K. Tunnel junctions were fabricated on 
monocrystalline
$\text{p-Hg}_{\text{1-x}}\text{Cd}_{\text{x}}\text{Te}$  with the
concentration of uncompensated  
acceptors $N_{A} -N_{D} = (2.5-60)\times10^{17}\text{ 
cm}^{\text{-3}}$.
The dopping level was  
determined from an analysis of galvanomagnetic phenomena in 
the temperature range of 1.5-70 K. Ultraviolet illumination 
for 5-15 minutes in dry air was used to form a thin oxide, which 
served as a tunnelling barrier. Then a metallic electrode (Yb) 
was evaporated through a mask. Several tunnel contacts prepared 
on each sample were investigated. The parameters of the 
tunnel structures are listed in the Table \ref{tab1}. It is
assumed that  the surface electric field in these structures
arises from the work  function difference of Yb and the
semiconductor. \cite{23} 

     The resistance of our structures is  0.1-1 kOhm and is 
determined mainly by the barrier transparency because the 
resistance associated with transition of electrons between 2D and 
bulk states of the semiconductor is significantly less in
metal-insulator-semiconductor structures based on inverted  
semiconductors. 

     The traditional modulation procedure was used for measuring 
the differential conductivity and its derivative $d\sigma/dV$. In some 
cases a refined method was used.\cite{23a} In parallel with a 
small (about 1 mV) alternative voltage with frequency $f=670$ Hz 
and direct bias $V$,  impulses with a period $T\gg1 /f$  and 
duration 
$t\ll 1/f$ were imposed across the tunnel contact (Fig.\ \ref{fig2}). 
These 
impulses lead to a change in the electric charge of the 
localized states which occur in the insulator or at the 
insulator-semiconductor boundary. If the relaxation time of these 
states far exceeds $T$, these impulses result in an increase or 
decrease 
(depending on sign of the impulses) of the strength of the 
surface potential well. Thus, measurements taken during 
the interval between impulses make it possible to get 
information about the energy spectrum of the 2D states with energy 
$E_{F}+eV$  but for a different strength of the surface potential 
well.

\section  {Experimental results}
\label{sec:exp}
     Let us consider first the results obtained from the measurements 
of the tunnel structures prepared on the most heavily doped 
sample (Table \ref{tab1}). The oscillation curves for such structures 
are 
simpler and therefore easier to interpret. Typical 
magnetic field dependences of the oscillatory part of the 
derivative of the tunnelling conductivity with respect to voltage 
at different biases are presented in Fig.\ \ref{fig3}.
Oscillations  are observed for both orientations of magnetic
field, $B \parallel n$ and $B \perp n$. The positions of the
maxima of $d^{2} j/dV^{2}$  for $B
\parallel n$ in the $B$, $V$ coordinates are plotted in Fig.\
\ref{fig4}. It is 
clear that with $B \parallel n$, every oscillation curve   
in Fig.\ \ref{fig3} is a convolution of several components and 
therefore it is 
difficult to follow the maxima positions of each component in 
Fig.\ \ref{fig4}. For  better distinction  a Fourier transformation of 
the oscillation curves is carried out. As shown in
Fig.\ \ref{fig5} for $B\perp n$ there is only one  
fundamental field  and for $B \parallel n$ two fundamental fields  
occur in the oscillation curves 
(a shoulder, which position coincides with the maximum position at
$B\perp n$, is resolved at some biases). Now one can 
separate  each component peak, take the inverse 
Fourier transform and follow the positions of the oscillation 
maxima of any one of the oscillation components independently. 
The curves in Fig.\ \ref{fig4} were obtained in this way and it can be
seen that they adequately describe all the experimental data. 

     As mentioned above (Eq.\  (\ref{eq1})) the period of 
oscillations $\Delta=B_{f}^{-1}$ is determined by the value of
quasimomentum. Bias  
dependences of $k$ for both orientations of magnetic field are 
plotted in Fig.\ \ref{fig6}. 

     For $B\perp n$, the spectrum of 2D states is not 
quntized by the magnetic field and, therefore, the oscillations of the 
tunnelling conductivity are due to tunnelling to Landau levels of 
bulk states only. Hence open circles in Fig.\ \ref{fig6} correspond to 
the 
energy spectrum of the bulk states.  The experimental data are
in  good agreement with $E(k)$ dependence calculated within the
Kane model with parameters: $P=8\times10^{-8} \text{eV cm}$,
$\Delta_{0}=\infty$ (where $\Delta_0$ is the energy difference 
between valence band $\Gamma_8$ and split-off $\Gamma_7$  
band)
and $E_{g}$  listed in the Table \ref{tab1}.  Extrapolation of this   
curve to $k=0$ gives the semiconductor Fermi energy $E_{F}
=-eV_{0}$ (Fig.\ {\ref{fig6}),
which is also given in the Table \ref{tab1}. 

     For $B \parallel n$, oscillations are connected mainly with 
tunnelling 
to 2D states localized in the surface quantum well of the 
semiconductor. This is evident from the typical angular dependence 
of the fundamental fields (Fig.\ \ref{fig7}). The reasons of some 
deviation from the classical angle dependence $B(\Theta)=B
(0)/cos(\Theta)$, where $\Theta$ is the angle between $B$ and
$n$,  were discussed in Ref.\ \onlinecite{19}. 
At $\Theta >20^{o}$  the $\Theta$-independent
maximum, which results from tunnelling to bulk Landau levels,
is resolved in the Fourier transform (open circles in Fig.\
\ref{fig7}).  Thus, Fig.\ \ref{fig6} shows that two branches of 2D  
states exist in this structure and at $k\rightarrow 0$ they are 
extrapolated to the bias $V_{0}>0$. This means that there are
no 2D states below  the 
Fermi level, i.e. 2D electrons are absent and therefore only 
oscillations of the first type (see Section \ref{sec:tun}) can be 
observed. 

    There are two possibilities for the explanation of these two 
branches; they are the ground and excited subbands of 2D states or 
they are two branches of the ground subband split by the spin-orbit 
interaction in the asymmetric quantum well. An estimation of the 
difference between the energies of the ground and excited subbands 
at $k=0$
shows that for structures based on a heavily doped
semiconductor with a small effective mass
(see Table \ref{tab1}) 
this value should be more than 50-100 meV (a
comparison will be given in more detal in Section \ref{sec:disc}).
This
value far exceeds the experimental one (Fig.\ \ref{fig6}). Thus two
branches of the 2D states  
in structure 10-1 (Fig.\ \ref{fig6}) correspond to two different spin 
branches of the ground subband which is split by spin-orbit 
interaction.
Hereafter supercripts + and - refer to different spin branches. 

     The structure 10-1 was also measured with voltage impulses 
which change the surface potential (see previous section) and
shift the 2D branches relative to the bulk states.
Fig.\ \ref{fig8}
shows this shift at V=80 mV versus the amplitude of the 
impulses. The distances between the branches of the 2D states and 
bulk states $\Delta(k^{+,-})^{2}=(k^{+,-})^{2}-
k_{\text{bulk}}^{2}$ 
versus quasimomentum of the lower branch of the 2D state 
$(k^{+})^2$
are plotted in Fig.\ \ref{fig9}. The results for the 10-2 to 10-6
structures, which differ 
by the value of the surface potential, are also shown in Fig.\ 
\ref{fig9}. 
It can be seen that when $(k^{+})^2$  decreases as the result of
reduction in the surface potential, the upper branch of the 2D states 
is pushed into the 
continuum, i.e. these localized 2D states disappear, and for
the 10-3 to 10-6 structures, in which
$\Delta(k^+)^2\leq 2\times10^{12}\text{ cm}^{-3}$, the only one
$k^+$-branch of 2D states exists.
The bias  dependence of the quasimomentum of 2D states 
for one of these structures (structure 10-4) over all bias
range are shown in  Fig.\ \ref{fig10}.

     The more complicated oscillation curves for $B \parallel n$ were 
observed 
for tunnel structures prepared on lesser doped samples. The 
maxima positions in $V$, $B$ coordinates  are plotted in Fig.\ 
\ref{fig11} for 
structure 12-1 . An inspection of Fig.\ \ref{fig11} shows that two 
main 
oscillation types are observed. The maxima of these oscillations 
shift in opposite directions relative to V as the magnetic field is 
varied. The behavior of the maxima positions with angle 
shows that both types of the oscillations are connected with 2D 
states. Taking the Fourier transform one can determine the 
fundamental fields, and using Eq.\ (\ref{eq1}) one can
calculate the quasimomentum of  
the states responsible for the oscillations at any bias (Fig.\ 
\ref{fig12}).
The oscillations for $B
\perp n$ as well as those for the structure 10-1  are due  
to tunnelling to the Landau levels of the bulk states. Thus the open
circles in Fig.\ \ref{fig12} are the energy spectrum of the
bulk states. The $1^+$, $0^-$ and $0^+$ branches correspond to
oscillations of the first  
type, i.e to oscillations arising whenever the Landau levels 
of the 2D states coincide with the energy $E_{F}+eV$. The branches
{\em a} and {\em b} relate to 
oscillations of the second type, i.e. to oscillations which 
arise whenever Landau levels of  2D states coincide with the energy 
$E_{F}$, therefore branches $0^+$, {\em b} and $0^-$, {\em
a} intersect at $V=0$ (Fig.\ \ref{fig12}). 
The decreasing of the quasimomentum at the energy $E=E_{F}$
with V (branches $a$, $b$) is a result of the decreasing
concentration  of 2D electrons due to a reduction of the depth
of the surface  potential well as bias is increased (inset in
Fig.\ \ref{fig12}).  We reasone that two branches of the ground 
2D subband, which is split by spin-orbit interaction, and one branch 
of the 
ecxited subband are observed for the structure 12-1 (Fig.\ 
\ref{fig12}).
Such interpretation will be confirmed in Section
\ref{sec:disc}. The analogous  
results were obtained for the 13-7 structure.

     It is significant that at negative biases up to $V=-30$ mV 
we observe  oscillations of the first type which occur 
whenever the Landau levels of  2D states coincide with the energy 
$E_{F}+eV$. 
At these energies the 2D states in the surface quantum well of 
the inverted semiconductor are in resonance with the 
valence band states. However in spite of this fact broadening
of the 2D states is not so large to destroy the oscillation
picture. To our knowledge this is the first experimental
observation of 2D states which lie significantly below the top
of the valence band.

\section{Theoretical model}
\label{sec:theor}

    To describe the energy spectrum of 2D system in
wide-gap semiconductors, the one-band approximation is
usually used. In this case the energy spectrum is
parabolic. Such an approach gives good results, because
typical energies of 2D states in wide-gap systems are
much less than the energy gap. In asymmetric quantum
systems (e.g., in surface quantum wells), as a result of 
spin-orbit interaction, splitting of the energy spectrum
arises even without an external magnetic field. It is
common practice to interpret the  experimental data in
this case using the Bychkov-Rashba model. \cite{9} The
spin-orbit interaction is described here by one additional
term, 
included in the dispersion law. It is linear in respect to
quasimomentum, and contains a new parameter, which has
to be calculated independently.

     The one-band model is inapplicable in the case of
2D states  in narrow gap and inverted semiconductors.
The strong interaction between the conduction and
valence band makes it necessary to employ a multiband
Hamiltonian in calculations of the energy spectrum of 2D
systems. It is well known that the energy spectrum of
$\text{A}^2\text{B}^6$ semiconductors can be described by the 
Kane Hamiltonian.
Because the Dirac Hamiltonian, which is simpler for calculations,
gives the same energy spectrum for electrons in the volume as the
Kane Hamiltonian,  the Dirac model is also widely used for
calculations of the energy spectrum of 2D systems in such
materials (see Refs.\ \onlinecite{11,24} and references
therein). However, this is not quite correct. It 
was demonstrated in Ref.\ \onlinecite{4} that parameters of the 
energy
spectrum of 2D electrons obtained with the Dirac and Kane
models, are radically different. The Kane model was used in our 
previous
article on the energy spectrum of 2D electrons in HgTe/HgCdTe
heterostructures.\cite{22} There are a number of points to be made
before applying this model in calculations of the energy
spectrum of 2D system near the interface oxide/semiconductor.

     There is difficulty in choosing the boundary
conditions at the  oxide/semiconductor interface, because
the energy band structure of oxide is unknown. The
interaction with remote bands is usually neglected in the
Kane model, when it is used for finding the spectrum of 2D
electron states. Then  the Schr{\"o}dinger equation is the system of
ordinary differential equations, and traditional boundary
conditions ($|\Psi|$=0 at the interface) lead in this
approximation to the existence of unique solution: the wave
function is equal to zero over all space.\cite{25}

     One way out of this dilemma is to reduce the system of  ordinary
differential equations to one equation of second
order which corresponds to one component of the wave function and 
to use
zero  boundary conditions only for this component.
\cite{10,26,27,28} This is a good  approximation for 2D electron
states in wide gap semiconductors, in which one
component of the wave function is much greater than the
other. However, all the components are of the same order of
magnitude in
narrow gap and inverted semiconductors, and additional
arguments are needed, as to why only one component
has to go to zero at the semiconductor/oxide interface.

    Another approach has been suggested by Sobkowicz in Ref.\ 
\onlinecite{25}. 
Here, the assumption is made that the band structure of the 
insulator is similar to that of the semiconductor. The only
difference is the value of the energy gap, which is much
greater for an insulator, than that for a semiconductor.
Moreover, the condition  $D_c,D_v\gg E$ (where $D_c$, $D_v$ are 
conduction and
valence band offsets, respectively) seems to be
natural, because neither electrons nor holes have to be 
emitted from semiconductor. Figure \ref{fig13} presents
schematically the model band structure for the case of
inverted semiconductors. Because the
electrostatic potential is constant at
$z<0$ and $z>L$, the exact wave functions are known in
these regions. In this case the eigenvalue problem can be solved 
exactly
using the techniques of direct numerical integration. \cite{22}

Thus in the framework of
this model one can understand, which insulator parameters 
correspond to the zero boundary condition for the second 
component of the wave function(see expressions (\ref{eq6}),
(\ref{eq8}) which follow) used in Ref.\ \onlinecite{28}. This can be
understood from Fig.\ 
\ref{fig14}, which shows the calculated dependence of the energy of  
the groung 2D subband on the value of $D_v$, while the value 
of $D_c$ is fixed (for details see Ref.\ \onlinecite{22}).  The
splitting of the 2D subband 
into two branches $0^{+}$ and $0^{-}$ results from spin-orbit
interaction. It is 
clearly seen, that zero boundary condition used in Ref.\ 
\onlinecite{28}
is a limiting case $D_v\rightarrow \infty$ and the limit is
reached very slowly. 
The results obtained in the above model come close to the 
solution of the zero boundary condition problem only at $D_v>100$ 
eV.

    The approach discussed above neglects the interaction with
remote bands which corresponds to infinite heavy hole mass. This
is suitable for calculation of the 
energy spectrum of 2D electrons only for positive
energies. When the energy of 2D states is negative,
they are in resonance with the continuous spectrum of
the heavy hole valence band. It is important to take remote bands 
into consideration in the calculation of
the energy spectrum in this energy range, because the
tunnelling of the carriers from the space charge layer
to the volume of a semiconductor can lead both to a 
change in the energy and a broadening of the 2D states.

    To calculate the energy spectrum of 2D states taking remote
bands into consideration, the usual
assumption has been made, that the energy difference $\Delta_0$
between valence band $\Gamma_8$ and split-off $\Gamma_7$  
band is very
large. In this case the Kane Hamiltonian is a $6\times6$
matrix. We choose 
the direction $z$ to be normal to the interface, and the
direction of carrier motion along $y$. Then the
Hamiltonian can be block diagonalized into two $3\times 3$
Hamiltonians for two groups of states. The Hamiltonian
matrix  for the first group is defined by
\widetext
\begin{equation}
H^+=
\left(
\begin{array}{ccc}
E^{\Gamma_6}+e\varphi(z) &
   i\sqrt{\frac{2}{3}}P\left(\frac{k_y}{2}-\frac{\partial}{\partial
   z}\right) & 
   \frac{i}{\sqrt{2}}Pk_y\\ \\
i\sqrt{\frac{2}{3}}P\left(-\frac{k_y}{2}-\frac{\partial}{\partial
   z}\right) & 
   \begin{array}{ccc}
\lefteqn{E^{\Gamma_8}+e\varphi(z)-} \\
&&-\frac{\hbar^2}{2m} \gamma_1 \left(k_{y}^{2}
-\frac{\partial^{2}}{\partial z^2}\right)
   \end{array}
 & 0\\ \\
-\frac{i}{\sqrt{2}}Pk_y & 0 &
   \begin{array}{ccc}
\lefteqn{E^{\Gamma_8}+e\varphi(z)-} \\
&&-\frac{\hbar^2}{2m} \gamma_1 \left(k_{y}^{2}
-\frac{\partial^{2}}{\partial z^2}\right)
   \end{array}\\
\end{array}
\right)
\label{eq5}
\end{equation}
\narrowtext where $E^{\Gamma_6}$, $E^{\Gamma_8}$ are the 
energies of
corresponding band edges, $\gamma_1$ is parameter, which
describes the interaction with the remote bands, and $\varphi(z)$ is
electrostatic potential. The values $P$ and $\gamma_1$ are assumed
to be the same for both the semiconductor and insulator. The
Hamiltonian $H^{-}$ for the second group of states is obtained from 
$H^{+}$ by
replacing $k_y$ by $-k_y$. Thus, the Schr{\"o}dinger equation
\begin{equation}
H^{\pm}\Psi=E^{\pm}\Psi
\label{eq5a}
\end{equation}
is a system of differential equations of second order,
which determines two branches of the energy spectrum
corresponding to two groups of states. The
eigenvectors in the insulator  and the volume of the
semiconductor  are known, because
$\varphi(z)=\text{const}$ at $z<0$ and $z>L$. There
are two types of eigenvectors. The first type
corresponds to a light particle and is
\begin{equation}
\Psi^{(l)}(k_y,k_z)=
\left(
\begin{array}{c}
E+\gamma_1\frac{\hbar^2 k^2}{2m}-E^{\Gamma_8} \\ \\
\sqrt{\frac{2}{3}}P\left(k_z-\frac{ik_y}{2}\right)\\ \\
\frac{i}{\sqrt{2}}Pk_y\\
\end{array}
\right)
e^{ik_y y+ik_z z}
\label{eq6}
\end{equation}
where the energy for fixed $k_{y}$, and $k_{z}$ is obtained from 
the
equation
\begin{equation}
\left(E-E^{\Gamma_6}\right)\left(E-
E^{\Gamma_8}+\frac{\gamma_1\hbar^2
}{2m}(k_{y}^{2}+k_{z}^{2})\right)=\frac{2}{3}P^2(k_{y}^{2}+k
_{z}^{2}). 
\label{eq7}
\end{equation}
The second type corresponds to
the heavy particle and is
\begin{equation}
\Psi^{(h)}(k_y,k_z)=
\left(
\begin{array}{c}
0 \\ \\
-\frac{i}{\sqrt{2}}Pk_y\\ \\
\sqrt{\frac{2}{3}}P\left(k_z+\frac{ik_y}{2}\right)\\ 
\end{array}
\right)
e^{ik_y y+ik_z z}
\label{eq8}
\end{equation}
for the energy
\begin{equation}
E=E^{\Gamma_8}+\frac{\gamma_1\hbar^2
}{2m}(k_{y}^{2}+k_{z}^{2})
\label{eq9}
\end{equation}
The wave function in the insulator $\Psi_{I}$ can be written as
linear combination of $\Psi^{(h)}$ and $\Psi^{(l)}$
\begin{equation}
\Psi_{I}=\Psi^{(h)}(k_y,-k^{(h)}_{z})+C\Psi^{(l)}(k_y,-
k^{(l)}_{z})
\label{eq10}
\end{equation}
where $C$ is some as yet unknown multiplier. Here only
the terms, which diminish at $z\rightarrow\infty$, are
given.

In the region $z>L$ the wave function is given by
\begin{eqnarray}
\Psi_{SC}&=&C_1\Psi^{(h)}(k_y,k^{(h)}_{z})
+C_2\Psi^{(l)}(k_y,k^{(l)}_{z})+ \nonumber \\
&& +C_3\Psi^{(h)}(k_y,-k^{(h)}_{z})+C_4\Psi^{(l)}(k_y,-
k^{(l)}_{z})
\label{eq11}
\end{eqnarray}
In Exps.\ (\ref{eq10}) and (\ref{eq11}) $k_{z}^{(l)}$, $k_{z}^{(h)}$ 
stand
for the quasimomentum 
components  perpendicular to the interface, which for any 
given $k_{y}$, and $E$ are determined from (\ref{eq7}) and 
(\ref{eq9}),
respectively. Because only normalised solutions are of
interest here, the coefficient $C_{4}$ in (\ref{eq11}) has to be zero.
So, we can now state the 
problem which can be numerically calculated: at a fixed energy
one needs to find a value of $C$ for $z<0$ such, that subsequent 
numerical integration of the Schr{\"o}dinger equation throughout
the region of the potential results in a zero value for $C_{4}$.

     The behaviour of the potential $\varphi(z)$ in
semiconductor is determined from the
Poisson equation for the charge density
\begin{eqnarray}
\rho(z)&=&-e|N_A-N_D|L\vartheta(L-z)-   \nonumber \\
&& -e\sum\int^{E_F}_{E_i}g(E)|\Psi(E,z)|^2
dE
\label{eq12}
\end{eqnarray}
where the summation runs over all occupied 2D subbands. $E_{i}$
denotes the energy of the bottom of $i$-th subband. The second term
describes the contribution of the electrons localized in
the quantum well. In the absence of 2D electrons 
(this situation occures in the 10-1 to 11-1 structures)
the second term in (\ref{eq12}) is equal to zero, and the Poisson
equation can be solved exactly. Then  $\varphi(z)$ is
parabolic:
\begin{equation}
\varphi(z)=
\left\{ 
\begin{array}{cc}
\varphi_s (1-z/L)^2, & 0\leq z \leq L\\
0 & z>L 
\end{array}
\right.
\label{eq13}
\end{equation}
where 
\[
\varphi_{s}=\varphi(0), 
\]
\[L=\left(\frac{2\kappa\kappa_0\varphi_s}{e(N_A-
N_D)}\right)^\frac{1}{2}
\]
and  $\kappa$ is the dielectric constant. For the structures with a 
concentration of 2D electrons comparable to $(N_A-N_D)\times L$
(as in structures 
12-1 and 13-7) one needs to calculate the potential self-consistently. 
We
use here the assumption made in Ref.\ \onlinecite{22}. Namely, at 
the
calculations of the charge density distribution we suppose that
the wave function is energy independent $\Psi(E,z)=\Psi(E_{F},z)$.

     Let us now consider peculiarities of the 2D states resulting from
resonance with the heavy hole band. All the results demonstrated in
this section has been obtained with the following parameter
values $E_g=-110$ meV, $P=8\times10^{-8}$ eV cm, 
$\gamma_1=2$,
$N_A-N_D=1\times10^{18}\text{ cm}^{-3}$, $\kappa=20$, and a 
parabolic dependence of $\varphi(z)$ with $\varphi_{s}=275$ mV. 
$D_c=2$ eV and $D_v=1$ eV are used (the choice of these values
will be justified in the next section). 

     Figure \ref{fig15} shows the $z$-dependence of $|\Psi(z)|^2$
for two energy 
values, corresponding to non-resonant ($E>0$) and
resonant ($E<0$)
2D electron states. It is clearly shown 
that the wave function does not decay in the semiconductor
region $(z>L)$ at negative energies. This means that the charge 
carriers from the space charge layer may go into the
volume of semiconductor, its wave function is transformed from
an electron into a heavy hole wave function in the process. To find 
the broadening of the 2D energy levels associated 
with such a resonance one may turn to the
scattering theory, namely Levinson's theorem.
\cite{25,29}. This allows a calculation of 
a density of states, added to the valence band by presence 
of the electrostatic potential $\varphi(z)$ and the
semiconductor/insulator interface, treated as a scattering potential:
\begin{equation}
g_{2D}\propto\frac{\int|\Psi(z)|^2dz}{A^2}
\label{eq14}
\end{equation}
where $A$ is the amplitude of $|\Psi|$ at $z>L$ 
and the integration runs over the space
charge layer region.

     Figure \ref{fig16} shows the density of states
corresponding to two groups of states, calculated for
different values of quasimomentum $k$ (in the following, $k$ means
$k_y$). It can be clearly seen that  
the width of maxima rapidly increases with decreasing 
quasimomentum values. Peculiar feature is that the width of the
maxima corresponding to the $0^{+}$ branch is greater, than that
of the  $0^{-}$
branch at the same energy. This means that tunnelling
of 2D electrons to the valence band states is more
effective for states of the $0^{+}$ branch. At $k=0$ the 2D energy 
levels
become well defined again.\cite{30}

     The dispersion law $E(k)$ of 2D electrons
calculated for the described model with above parameters is
presented in Fig.\ \ref{fig17}. At $E>0$ the results coinside
with those obtained without considering the remote bands. 
At  negative
energies when the 2D states are resonant, the energy levels are
broad. This broadening is shown in Fig.\ \ref{fig17} as hatched 
regions. The  
width of these regions is defined here as the half-width 
of the peak of the density   
of 2D states (see Fig.\ \ref{fig16}). In the narrow range of
$k^2=(0.3-3)\times10^{12}\text{ cm}^{-2}$, the width of
the density of states peak is very large, i.e. the 2D states are
practically destroyed in this range.

     Earlier, the resonance broadening of 2D states was discussed in
Ref.\ \onlinecite{25}   for narrow gap semiconductors and in
Ref.\ \onlinecite{28}  for 
semiconductors with inverted spectrum. In these articles the
tunnelling to light hole states was taken into account, but 
tunnelling to heave hole states was neglected.  In such
approximation the broadening of 2D states in inverted
semiconductors occures only at the energy less than 
$E^{\Gamma_6}$ in 
the bulk of semiconductor. As shown above the tunnelling to the
heavy hole states is effective and this process has to be
taken into account in inverted semiconductors. 

\section{Discussion}
\label{sec:disc}

     Let us begin with the discussion of experimental results 
obtained for the tunnelling structures 10-1 to 11-1, based on heavily 
doped materials.

     As  mentioned above there are two ways to 
understand the presence of two branches in the energy spectrum
of the 2D states in the 10-1 
structure. They are either ground and excited 2D subbands 
or two spin-split states of the ground 2D subband. The calculations 
carried out in the framework of the above model show (see Fig.\
\ref{fig18}) that  
as the surface potential increases, the excited subband at $k=0$
appears only when the energy of the ground subband is less 
than -100 meV (this value only slightly depends on the values of 
$D_c$ and $D_v$). This is in contradiction with the experimental 
results 
presented in Fig.\ \ref{fig6}. Indeed, extrapolating the experimental 
curves 
to $k=0$ resuits in an energy difference between these brunches 
which
does not exceed 10 meV. Thus, these two branches of the 2D
states are two spin branches of the 2D subband split by
spin-orbit interaction.  

     For quantitative analysis of the experimental data the 
parameters $D_c$, $D_v$, and $\varphi_{s}(V)$ relationship
should first be determined. The experimental 
results for the structures 10-1 to 10-6 provide a way of estimating 
the parameters $D_c$ and $D_v$  independent of
$\varphi_{s}(V)$. In reality, we can  
calculate the $\varphi_{s}$ relationships of the quasimomentum for
the upper 
$k^-(\varphi_{s})$ and lower $k^+(\varphi_{s})$ branches of the 2D 
states
at a fixed energy.  
Then eliminating $\varphi_{s}$ we obtain a $k^-$ versus
$k^+$ dependence which is  
shown in the $\Delta (k^{+,-})^2$, $(k^+)^2$ coordinates in Fig.\
\ref{fig9} for different values of $D_c$ and 
$D_v$. The inspection of Fig.\ \ref{fig9} shows that the results are
not very sensitive to $D_c$ and $D_v$  as  
long as $D_c$ and $D_v>E_g$  and $D_c \approx D_v$. But when
$D_v>D_c$  and especially  
in the case when $D_v=\infty$, which corresponds to the zero 
boundary 
condition for the second component of the wave function, the 
results of calculation significantly deviate from the experimental
results. On the first sight it seems to be surprising that the authors
of Ref.\ \onlinecite{28}   were able to explain the
experimental results on the spin-orbit splitting of the 2D
energy spectrum using such a type of the boundary condition. The 
analysis of
this article shows that this is due to using of the
approximate (quasiclassical) method of the solving of the
eigenvalue problem in Ref.\ \onlinecite{28}. Indead, as is seen
from Fig.\ \ref{fig9} the results of the approximate calculations
(dashed line) differ significantly from the exact solutions (line
4), but they are in a better agreement with the experimental
data. 

Fig. \ref{fig9} shows that the more suitable parameters are
$D_c=2$ eV, $D_v =1$ eV, and they will be used for the analysis
of the results for all tunnel structures.  

     We have calculated the bias dependence of the 
quasimomentum of the lower spin branch over  the entire bias range 
for 
structure 10-1, using the surface potential $\varphi_{s}$  as a
fitting parameter  
(inset in Fig.\ \ref{fig6}). With this $\varphi_s(V)$
relationship we have calculated the position of the 
upper spin branch of the 2D states (Fig.\ \ref{fig6}). One can see the 
good 
agreement with spin splitting over the whole energy range. Notice 
that at 
$k=0$ the energy of the 2D states coincides with the energy of the 
bottom of the conduction band, i.e. the binding energy of the 
2D states at $k=0$ is equal to zero. This is a specific feature of 
the 2D states in an inverted semiconductor for weak quantum
well strengths.

     Curve 1 in Fig.\ \ref{fig9} shows that the upper spin
states must disappear,  when $\Delta (k^+)^2$  for the
lower  spin branch is less than $4\times10^{12}\text{
cm}^{-2}$, which corresponds to a small surface potential
$\varphi_s<230$ meV. Actually for the  
10-3 to 10-6 structures only one spin branch exists over the whole 
energy range. This is illustrated by Fig.\ \ref{fig10}, which shows the 
energy 
versus quasimomentum curve for one of these strictures (10-4). It 
can be clearly seen that the model employed describes the 
experimental 
data for these structures well.

    Let us inspect the results for the structure 12-1 
(Fig.\ \ref{fig12}). Contrary to structures of the first type
(10-1 to 11-1) there are 2D electrons in this structure. Their
concentration can be determined from the bias dependences of
$k^2$ for oscillations of the second type (branches {\em a} and
{\em b} in Fig.\ \ref{fig12}).
The concentration is varied from $n_{2D}=1.3\times
10^{12}\text{ cm}^{-2}$ at $V=-60$ mV to  $n_{2D}=1.0\times
10^{12}\text{ cm}^{-2}$ at $V=50$ mV.
An estimation of the density of uncompensated acceptors in the 
space charge region gives a value of about
$2\times10^{12}\text{ cm}^{-2}$, which is of  
the same order of magnitude as $n_{2D}$, therefore the self-
consistency of 
the potential has to be included in the calculation.  
Knowing the density of 2D electrons at different biases (see 
section \ref{sec:exp}) and using a self-consistent procedure we have 
calculated the bias dependence of the surface potential (inset 
in Fig.\ \ref{fig12}). One can see that $\varphi_{s}$ for the
structure 12-1 is close  
to that of the structure 10-1 but due to the lower doping level (Table 
\ref{tab1}) the width of the potential well is significantly larger 
so that the strength of the potential well is also larger and more 
than one 2D subbands are localized in such a well. 
Using this relationship between $\varphi_{s}$ and bias we have 
calculated the 
quasimomenta of 2D states over the entire energy range (curves
$1^+$, $0^-$ and $0^+$ in Fig.\ \ref{fig12}). Curves {\em a}
and {\em b} in Fig.\  
\ref{fig12} present the calculated values 
of the quasimomenta for both spin states at the energy 
$E=E_F$. As can be seen, there is good agreement for both types of
oscillations over the whole bias range. Thus, both spin
branches of the  ground 2D subband and lower spin branch of the
excited subband  exist in the surface quantum well of this structure.

     As was mentioned above (section \ref{sec:theor}), the 2D states 
should 
broaden at negative energy due to resonance with the valence 
band. As well as in Fig.\ \ref{fig17} this broadening is shown
in Fig.\ \ref{fig12} by
hatched regions. One can see that at the bias, where 
oscillations were observed  experimentally, broadening does
not exceed 5 meV, i.e. it is less  than the cyclotron energy in the
investigated semiconductors at  the magnetic fields, where
oscillations were detected ($B > 1$ T). With a more negative bias,
the broadening becomes significantly  larger and this is
conceivable reason why we do not observe  oscillations at
these biases. 

    Two comments are necessary in closing this section. 

    At first sight it may seem that the model of the insulator used 
for calculations of the energy spectrum of 2D states in our 
structures is  forced. What actually happens is that 
this model imposes some relationships between the 
components of the wave function at the interface. Any insulator 
considered in the framework of the kP-method imposes
some relationships between the components too.  
The only point is then what relationships does it impose. In 
inverted and narrow gap semiconductors the amplitudes of all the 
components are about of the same value, therefore a widely used zero 
boundary
condition for one of the component of the wave function 
\cite{10,26,27,28}
seems to be artificial.   

     All the experimental results on the energy spectrum of 2D 
states have been obtained from an analysis of oscillations of 
tunnelling conductivity in a magnetic field, but the 
calculations were carried out without a magnetic field. For the 
comparison of the calculated results with the experimental data, 
the assumption was used that the quasiclassical rule for the 
quantization of the energy spectrum of 2D states in a magnetic field 
was fulfilled. To check this assumption, calculations of
the Landau levels were also carried out. It was shown that
except for the low surface potential values, when the energy
difference  between
2D and bulk states with the same quasimomentum was small, the
quasiclassical rule for quantization in magnetic fields was
fulfilled with good accuracy. Some results of these calculations
and analysis of the peculiarities of the behavior of 2D Landau
levels were discussed in Ref.\ \onlinecite{19}. 

\section{Conclusion}
\label{sec:conc}
     The energy spectrum of 2D states localized in a surface quantum 
well in inverted semiconductor was studied by tunnelling 
spectroscopy in a quantizing magnetic field. In the investigated 
structures 
the strength of potential well mainly depends on the 
width of the well which is controlled by the doping level: it
increases when the acceptor concentration decreases. Two
different cases were  observed in our structures. 

     In the structures based on heavily doped material only
ground 2D subband exists and its bottom coincides with the bottom of 
the conduction band. Thus, in p-type material there are no 2D 
electrons and only empty 2D states exist. At $k\neq 0$ the 2D ground 
subband is split by spin-orbit interaction. In 
structures with a larger potential well strength,  both spin 
branches are observed, whereas in other structures the upper spin 
states are pushed into the continum and only the lower spin branch is 
observed. 

     In the structures based on materials with
$N_A-N_D<10^{18}\text{ cm}^{-3}$
the strength of the potential well is so large that the ground 
and excited 2D subbands exist in these structures. In this case 
both spin branches of the ground subband and only lower spin
branch of the excited subband are  observed. In these
structures the bottom of the ground subband  
lies significantly lower than the bottom of the conduction band 
and the 2D states are in resonance with the valence band over a
wide energy range. The fact that oscillations relating to 
tunnelling to such states are observed, shows that the 
resonance broadening is not large and the 2D states are not
destroyed at these energies. 

     Theoretical calculations carried out in the framework 
of the Kane model describe well all the peculiarities 
of the energy spectrum and spin-orbit splitting of 2D states in 
inverted semiconductors.
Theoretical calculations which take 
the finite value of the heavy hole effective mass into account
have shown that the broadening of 2D  
states due to resonance with the heavy hole valence band is 
large only over a narrow range of quasimomentum values. This is 
consistent with the results of our tunnelling experiments.

\acknowledgments

The authors would like
to thank Dr.C.R.Becker for critically reading the manuscript.
This work was supported, in part, by the programme {\em 
Universities
of Russia} and by a grant from the State Committee of the
Russian Federation on Higher Education.

\begin{figure}
\caption {Energy diagram of a metal (M)-insulator-inverted
semiconductor (SC)
tunnel junction with a surface quantum well for a bias $V>0$. } 
\label{fig1}
\end{figure}

\begin{figure}
\caption{Voltage which is applied to a
tunnel junction.  The voltage is a sum of a dc bias
$V$, an ac modulation voltage with amplitude $V_{m}$, and 
impulses
with amplitude $V_{i}$. }
\label{fig2}
\end{figure}

\begin{figure}
\caption{$d^2j/dV^2$ versus $B$ curves at different biases for 
for the structure 10-1. }
\label{fig3}
\end{figure}

\begin{figure}
\caption{The positions of the maxima of the $d^2j/dV^2$ versus $B$ 
curves 
for the structure 10-1 for  $B\parallel n$.  The curves are
obtained in the way described in 
the text. } 
\label{fig4}
\end{figure}

\begin{figure}
\caption{The results of Fourier transformation of the tunnelling
curves plotted in Fig. \ \protect\ref{fig3}. A shoulder on a low-field 
peak
for a bias of 100 mV for $B\parallel n$ corresponds to tunnelling to 
the bulk
Landau levels. }
\label{fig5}
\end{figure}

\begin{figure}
\caption{Bias dependences of $k^2$ for the structure 10-1.  
Open and full circles represent 
the experimental data for $B\perp n$ and $B\parallel n$,
respectively.  The upper curve   
is the dispersion law of the bulk electrons, calculated in the
framework of the Kane model with parameters $E_g=-70$ meV,
$P=8\times 10^{-8}$ eV cm.  The other curves 
are the result of theoretical calculations for the 2D states
described in Section \protect\ref{sec:theor}.  
Inset shows the bias
dependence of $\varphi_s$. }
\label{fig6}
\end{figure}

\begin{figure}
\caption{Dependence of fundamental magnetic fields for
the structure 10-1, measured at $V=80$ mV.  Dashed line
indicates  the $cos(\Theta)^{-1}$ dependence.  Open and full circles
relate to tunnelling to the bulk and 2D states, respectively. }
\label{fig7}
\end{figure}

\begin{figure} 
\caption{A quasimomentum of 2D states as a function of the
amplitude of impulses for a bias of $V=80$ mV, which corresponds
to the energy $E=E_{F}+eV=65$ meV, for the structure 10-1.  
Dashed line shows
the value of $k^2$ of the bulk electrons for the same energy. 
Solid curves are merely guide for the eye. }
\label{fig8}
\end{figure}

\begin{figure}
\caption{The differences
$\Delta(k^+)^2$ and $\Delta(k^-)^2$ as a
function of $(k^+)^2$ for different structures, $V=80$ mV.  
Points for structure 10-1 have been obtained at different
amplitudes of impulses.  Solid curves are the result of
numerical calculations carried out with parameters listed in
the table and different band offset values: $D_{c}=2$ eV, 
$D_{v}=1$ eV
for curve 1, $D_{c}=5$ eV, $D_{v}=1$ eV for curve 2, $D_{c}=1$
eV, $D_{v}=1$ eV 
for curve 3.  The curve 4 and dashed curve are the results of
exact and approximate (in accordance with Ref. \ 
\protect\onlinecite{28})
calculations with zero boundary condition for 
the second component of the wave function.  The upper curve is
the same for the calculations with different parameters. }
\label{fig9}
\end{figure}

\begin{figure}
\caption{Bias dependences of $k^2$ for the structure 10-4.  
Open and full circles represent 
the experimental data for $B\perp n$ and $B\parallel
n$, respectively.  The upper curve  
is the dispersion law of the bulk electrons, calculated in the
framework of the Kane model with parameters $E_g=-70$ meV,
$P=8\times10^{-8}$ eV cm.  The other curve 
is the result of theoretical calculations for the 2D states
described in Section \protect\ref{sec:theor}.  Inset shows the bias
dependence of $\varphi_s$. }
\label{fig10}
\end{figure}

\begin{figure} 
\caption{Fan chart diagram for structure 12-1 at  $B\parallel n$
orientation.  The solid curves are merely guide for the eye. } 
\label{fig11}
\end{figure}

\begin{figure}
\caption{The bias dependences of $k^2$ for structure 12-1.  Symbols
show the experimental data: the open and full circles correspond to 
the bulk
and 2D states respectively.  The dotted curve is the dispersion
law of the bulk electrons, calculated in the framework of the
Kane model with parameters
$E_{g}=-110$ meV, $P=8\times10^{-8}$ eV cm.  The solid curves 
are the
theoretical bias dependences of $k^2(E_{F}+eV)$ (curves $1^+$, 
$0^+$,
$0^-$) and $k^2(E_{F})$ (curves {\em a}, and {\em b}).  Inset shows 
the bias
dependence of $\varphi_s$.  The hatched regions show the 
broadening of
the 2D energy levels. }
\label{fig12}
\end{figure}

\begin{figure}
\caption{Model of an insulator-inverted semiconductor structure
with a surface quantum well used in the calculations. }
\label{fig13}
\end{figure}

\begin{figure}
\caption{The dependence of the energy of two branches of the 2D
ground subband at $k=3\times10^{6}\text{ cm}^{-1}$ on the 
valence band offset
value while $D_c$ is kept constant $D_c=1$ eV.  The dashed lines 
are
the energies of the branches calculated under zero boundary
condition for the second component of the wave function at the 
insulator/inverted semiconductor interface.  The calculations
have been carried out with parameters $E_{g}=-70$ meV,
$P=8\times10^{-8}$ eV cm, $N_A-N_D=6\times10^{18}\text{
cm}^{-3}$, $\varphi_s=240$ meV, and for parabolic surface
quantum well. }   
\label{fig14}
\end{figure}

\begin{figure}
\caption{$|\Psi|^2$ versus {\em z} curves for 
resonant (lower figure) and nonresonant (upper figure) 2D states.  The
parameters used are in the text. }
\label{fig15}
\end{figure}

\begin{figure}
\caption{The density of resonant 2D states
for different quasimomentum values for $0^+$ (a) and $0^-$ (b) 
branches. }
\label{fig16}
\end{figure}

\begin{figure}
\caption{The dispersion law of the 2D states localized in a surface
quantum well.  Two subbands split by spin-orbit interaction are
shown.   The hatched regions show the broadening of the 2D
energy levels.  The 2D states of the $0^{+}$ 
branch are distroied in the range $k^2=(0. 3-3. 0)\times10^{12} 
\text{ cm}^{-2}$ due to high probability of tunnelling into the heavy 
hole
states of the volume of inverted semiconductor. }
\label{fig17}
\end{figure}

\begin{figure}
\caption{The energy of the bottoms of the ground and first
excited 2D subbands as a function of surface quantum well depth,
calculated with parameters of structure 10-1. }
\label{fig18}
\end{figure}

\newpage
\mediumtext
\begin{table}
\caption{Parameters of investigated structures. 
\label{tab1}}
\begin{tabular}{cccccccc}
structure & x & $E_g$ (meV) &$m_n\,(10^2m)$\tablenotemark[1] &
$N_A-N_D\,(10^{17}\text{cm}^{-3})$ & $E_F$ 
(meV) & $\varphi_s\,$(meV)\tablenotemark[2] &
$L\,(10^{-6}\,\text{cm})$\tablenotemark[2] \\ 
\tableline
10-1 & & & & & &$275\pm5$&1.0\\
10-2 & & & & & &$270\pm5$&1.0 \\
10-3 &0.125 &$-70\pm5$ &6.25 &60 &$-15\pm5$ &$240\pm5$&0.95 \\
10-4 & & & & & &$230\pm5$&0.92\\
10-5 & & & & & &$210\pm5$&0.88\\
10-6 & & & & & &$180\pm5$&0.8\\
11-1 &0.115 &$-90\pm5$ &8.0 &20 &$-10\pm5$ &$200\pm5$ &1.5\\
12-1 &0.105 &$-110\pm5$ &9.75 &8 &$-8\pm3$ &$275\pm5$ &2.75\\
13-7 &0.095 &$-125\pm5$ &11.5 &5 &$-5\pm2$ &$260\pm5$  &2.85 \\
\end{tabular}
\tablenotetext[1]{The effective mass of electrons at 
the conduction band bottom in the volume of semiconductor, $m$
is the free electron mass. }
\tablenotetext[2]{The values are given for $V=80$
mV for the 10-1 to 11-1 structures, and for $V=0$ mV for other 
structures. }
\end{table}
\narrowtext

\end{document}